\documentstyle[11pt,aaspp4]{article}

\slugcomment{\shortstack[r]{
Submitted to the ApJ, 1998 January 14}}
\def\etal{{\it et al.\ }}
\def\Order{{\cal O}}
\def\pbf{{\bf p}}
\begin{document}

\title{Hot, Rotating Disks In General Relativity: 
                   Collisionless Equilibrium Models}
\author{A. Katrin Schenk \altaffilmark{1}, Stuart L. Shapiro \altaffilmark{2,3}, and Saul A. Teukolsky \altaffilmark{1,4}}
\altaffiltext{1}{Center for Radiophysics and Space Research, Cornell
 University, Ithaca, NY 14853}

\altaffiltext{2}{Department of Physics, University of
Illinois at Urbana-Champaign}
\altaffiltext{3}{Department of Astronomy and
NCSA, University of Illinois at Urbana-Champaign, Urbana, IL 61801}
\altaffiltext{4}{Departments of
Physics and Astronomy, Cornell University}
\date{\today}

\begin{abstract}
We present a method for constructing equilibrium disks with net 
angular momentum in general relativity.  The method solves the relativistic
Vlasov equation coupled to Einstein's equations for the gravitational field.  We apply the method to construct disks that are relativistic versions of 
Newtonian Kalnajs disks.  In Newtonian gravity these disks are analytic,
and are stable against ring formation for certain ranges of their velocity
dispersion. We investigate the existence of fully general relativistic 
equilibrium sequences for differing values of the velocity dispersion.
These models are the first rotating, relativistic disk solutions of the 
collisionless Boltzman equation.

\end{abstract}

\section{\bf INTRODUCTION}\label{intro}

Rotating stellar disks in dynamical equilibrium have a long history
in astrophysics.  Such systems are described by self-consistent
solutions to the Vlasov equation for the phase-space distribution 
function $f$ coupled to the equations for the gravitational field.
Even in Newtonian gravitation, finding solutions is difficult since the 
configurations are nonspherical and have a relatively large
number of nontrivial phase space degrees of freedom.  (For a 
general review and discussion, see \cite{frid}
or \cite{bnt}.)  When the configuration is relativistic, the 
gravitational field is described by Einstein's equations and hence the
problem is even more difficult.

In a previous paper (Shapiro \&  Teukolsky 1993; hereafter \cite{st1}), we developed a 
method for constructing equilibrium axisymmetric star clusters with
net rotation in general relativity.  In this paper we adapt this method
to treat disks. Once again we restrict our attention to the simplest 
phase-space distribution functions that can give rise to nonspherical
equilibria, functions of particle energy $E$ and angular momentum $J_z$
alone.  Because $E$ and $J_z$ are integrals of the motion, choosing a 
distribution function of the form $f=f(E,J_z)$ guarantees that we 
will have  a solution to the Vlasov equation, provided the metric is
determined self-consistently.  No further dynamical equations need to be 
solved for the matter.  By contrast, in equilibrium fluid systems one must 
integrate the equation of hydrostatic equilibrium.

Several researchers have studied the properties of self-gravitating
{\it fluid} disks in general relativity.  \cite{olddisk1()} presented an analytic model
of a fluid disk with no net angular momentum. This work was followed
by \cite{olddisk2()} who studied rapidly rotating, semi-analytic, 
zero pressure models and by \cite{olddisk3()} who looked at rapidly
rotating fluid disks with some thickness.  Recently the model presented
by \cite{olddisk2()} was solved analytically by \cite{anfluiddisk()}.
Collisionless disks with no net angular momentum have been
modeled and evolved using a particle simulation code by \cite{nocolldisk}.  
The
work presented here presents the first fully general relativistic,
rotating, collisionless disk models.

As an illustration of our method, we will focus on the relativistic 
generalization of an important class of Newtonian disks, the Kalnajs disks
(\cite{kal}; \cite{bnt}).  These disks are completely described by simple
analytic expressions.  In the limiting case that all the particles move
in circular orbits, the angular velocity and surface density are just
those of the corresponding fluid Maclaurin spheroid in the disk limit
(eccentricity $\rightarrow 1$).  In Newtonian theory, however, it
is known that equilibrium disks supported against collapse by rotation alone
are unstable to ring formation (see, {\it e.g.}, \cite{bnt}, \S5.3).
\cite{kal()} showed that the disk can be stabilized by ``heating'' it,
that is, converting some of the ordered rotational energy into random
thermal motion while keeping the surface density the same.  (Properties
of Newtonian Kalnajs disks are reviewed in Appendix A.)  We construct 
relativistic generalizations of Kalnajs disks and study their properties.

In addition to its potential astrophysical significance for generating
models of highly relativistic stellar disks or collisionless particle
distributions, the method presented here provides a class of rotating
equilibria that has not been treated previously in the general
relativistic literature.  These collisionless models join Kerr black holes and 
rotating fluid stars and fluid disks as physically realistic rotating 
equilibria in general relativity.  Such solutions to Einstein's equations
provide important insight into the effects of rotation in a strong
gravitational field.  They also provide useful initial data for evolution
codes in general relativity.  In particular, such codes can assess the 
stability of relativistic Kalnajs disks, since, as we will see below,
we can only make heuristic statements about their stability from the results
of this paper. 

\section{\bf Basic Equations}\label{basiceqs}
In this paper we consider rotating
equilibrium stellar disks that are axisymmetric.
The metric can then be written in the form
\begin{equation}
ds^2 = - e^{\gamma + \rho}dt^2 + e^{2\sigma}(dr^2 + r^2d\theta^2) +
        e^{\gamma - \rho}r^2\sin^2\theta (d\phi - \omega dt)^2
\end{equation}
where the metric potentials $\rho$, $\gamma$, $\omega$, and $\sigma$ are
functions of $r$ and $\theta$ only.  Here and throughout we set
$G=c=1$.  This is the same form of the metric used in \cite{st1}.

All calculations in this paper are performed in the ZAMO frame (\cite{bar}; 
 \cite{zamo}). This is the orthonormal frame of a zero angular
momentum observer, whose basis 1-forms $\omega^{\hat\alpha}$
are related to the coordinate basis $dx^\alpha$ by
\begin{equation}
	\omega^{\hat\alpha} = L^{\hat\alpha}{}_{\alpha} dx^{\alpha}
\end{equation}
where
\begin{equation}
	L^{\hat\alpha}{}_{\alpha} = \left[ \matrix{e^{\nu} &0 &0 &0\cr
	0 &e^{\nu} &0 &0\cr
	0 &0 &re^{\nu} &0\cr
	- \omega r \sin \theta e^{\beta} &0 &0 &r 
	\sin \theta e^{\beta}\cr} \right]
\end{equation}
and where
\begin{eqnarray}
	\nu &=&{\gamma+\rho\over 2},\\
	\beta&=&{\gamma-\rho\over 2}.
\end{eqnarray}
The corresponding orthonormal basis vectors are given by
\begin{equation}
	{\vec e_{\hat \alpha}} = L^{\alpha}{}_{\hat\alpha} {\partial
	\over \partial x^{\alpha}}
\end{equation}
where $L^{\alpha}{}_{\hat\alpha}$ is the inverse matrix of
$L^{\hat\alpha}{}_{\alpha}$.

Following \cite{koma}, \cite{cook()}, and \cite{st1}, we write the Einstein field equations that
determine $\rho$, $\gamma$ and $\omega$ in the form
\begin{equation}
	\nabla^2[\rho e^{\gamma /2}] = S_\rho(r,\mu),\label{eq:ein1}
\end{equation}
\begin{equation}
	\left(\nabla^2 + {1\over r}\partial_r - 
	{\mu\over r^2}\partial_\mu \right)
        [\gamma e^{\gamma / 2}] = S_\gamma(r,\mu),\label{eq:ein2}
\end{equation}
\begin{equation}
	\left(\nabla^2 + {2\over r}\partial_r - 
	{2\mu\over r^2}\partial_\mu \right)
        [\omega  e^{(\gamma - 2\rho)/ 2}] = S_\omega(r,\mu),\label{eq:ein3}
\end{equation}
where $\nabla^2$ is the flat-space, spherical coordinate scalar Laplacian,
$\mu = \cos\theta$, and $S_\rho$, $S_\gamma$ and $S_\omega$ are effective
source terms that include the nonlinear and matter terms.  In the equations
below, we explicitly exhibit the disk nature of the matter by defining 
a disk stress-stress energy tensor, $t^{\hat \mu}_{\hat \nu}(r)$ such that
the full stress-energy tensor, $T^{\hat \mu}_{\hat \nu}(r,\mu)$ is given by
\begin{equation}
T^{\hat \mu}_{\hat \nu}(r,\mu) = t^{\hat \mu}_{\hat \nu}(r){\delta (\mu) \over r}.\label{fullSE}
\end{equation}
 Thus  the integrals used to calculate
the stress energy components, equations~ (\ref{eq:t1})-(\ref{eq:t03}) below,
need only  be evaluated in the disk plane ($\mu
=0$).

The effective
source terms are given by
\begin{equation}
S_\rho(r,\mu) \equiv R_\rho(r,\mu) + t_\rho(r){\delta (\mu) \over r}
\end{equation}
\begin{equation}
S_\gamma(r,\mu) \equiv R_\gamma(r,\mu) + t_\gamma(r){\delta (\mu) \over r}
\end{equation}
\begin{equation}
S_\omega(r,\mu) \equiv R_\omega(r,\mu) + t_\omega(r){\delta (\mu) \over r}
\end{equation}
where we have defined matter dependent and matter independent source
terms such that
\begin{eqnarray}
R_\rho(r,\mu) &=& e^{\gamma / 2}\Biggl\{
        {1\over r}\gamma_{,r} - {\mu\over r^2}\gamma_{,\mu} +
        {\rho\over 2}\left[ -
        \gamma_{,r}\left({1\over 2}\gamma_{,r} + {1\over r}\right) -
        {1\over r^2}\gamma_{,\mu}\left({1 - \mu^2\over 2}\gamma_{,\mu} -
        \mu\right) \right] \nonumber \\
        & & + r^2(1 - \mu^2)e^{-2\rho}
        \left(\omega_{,r}^2 + {1 - \mu^2 \over r^2}\omega_
	{,\mu}^2\right)\Biggr\}\\
t_\rho(r,\mu) &=&8\pi  e^{\gamma / 2}e^{2\sigma}
	\left[t^{\hat \phi}_{\hat \phi}
        -t^{\hat t}_{\hat t}
        + {\rho\over 2}(t^{\hat r}_{\hat r}+t^{\hat \theta}_{\hat \theta})
        \right],\label{eq:trho}\\
R_\gamma(r,\mu) &=& e^{\gamma / 2}
        {\gamma\over 2}\left[  - {1\over 2}\gamma_{,r}^2 -
        {1 - \mu^2 \over 2r^2}\gamma_{,\mu}^2 \right],\\
t_\gamma(r,\mu) &=&8\pi e^{\gamma / 2}
	e^{2\sigma}\left(1+{\gamma\over 2}\right)(
         t^{\hat r}_{\hat r}+t^{\hat \theta}_{\hat \theta})
	,\label{eq:tgam}\\         
R_\omega(r,\mu) &=& e^{(\gamma - 2\rho)/ 2}
        \omega\left[
        -{1\over r}\left(2\rho_{,r} + {1\over 2}\gamma_{,r}\right) +
        {\mu\over r^2}\left(2\rho_{,\mu} + {1 \over 2}\gamma_{,\mu}\right) +
        {1\over 4}(4\rho_{,r}^2 - \gamma_{,r}^2)\right.\nonumber \\
        & &\left. +{1 - \mu^2\over 4r^2}(4\rho_{,\mu}^2 -
        \gamma_{,\mu}^2)
        - r^2(1 - \mu^2)e^{-2\rho}\left(\omega_{,r}^2 +
        {1 - \mu^2\over r^2}\omega_{,\mu}^2\right)\right],\\
t_\omega(r,\mu) &=&  8\pi e^{(\gamma - 2\rho)/ 2}
	e^{2\sigma}\Biggl[-\omega(t^{\hat \phi}
	_{\hat \phi}
        -t^{\hat t}_{\hat t})
        + {\omega\over 2}(t^{\hat r}_{\hat r}+t^{\hat \theta}_{\hat \theta})
        -{2e^\rho t^{\hat t}_{\hat \phi}\over r (1-\mu^2)^{1/2}}
        \Biggr]
        \label{eq:tomg}. 
\end{eqnarray}
Here $t^{\hat \mu}_{\hat \nu}$ are the orthonormal components of our disk 
stress-energy tensor for collisionless matter in the ZAMO frame (see below).

The fourth field equation determines $\sigma$ and is given by
\begin{eqnarray}
\sigma_{,\mu} &=& -{1\over 2}(\rho_{,\mu} + \gamma_{,\mu}) -
        \{(1 - \mu^2)(1 + r\gamma_{,r})^2 +
                [\mu - (1 - \mu^2)\gamma_{,\mu}]^2\}^{-1}\nonumber \\ 
        & &\times\Biggl\{{1\over 2}[r^2(\gamma_{,rr} + \gamma_{,r}^2) -
        (1 - \mu^2)(\gamma_{,\mu\mu} +
                \gamma_{,\mu}^2)][-\mu + (1 - \mu^2)\gamma_{,\mu}] \nonumber \\
         & & + r\gamma_{,r}\left[{1\over 2}\mu + \mu r\gamma_{,r} +
                {1\over 2}(1 - \mu^2)\gamma_{,\mu}\right] +
                {3\over 2}\gamma_{,\mu}[-\mu^2 + \mu(1 - \mu^2)
                \gamma_{,\mu} ]\nonumber \\
         & & - r(1 - \mu^2)(\gamma_{,r\mu} + \gamma_{,r}\gamma_{,\mu})
                (1 + r\gamma_{,r}) - {1\over 4}\mu r^2(\rho_{,r} +
                \gamma_{,r})^2\nonumber \\
        & & - {r\over 2}(1 - \mu^2)(\rho_{,r} + \gamma_{,r})
                (\rho_{,\mu} + \gamma_{,\mu}) + {1\over 4}\mu(1 - \mu^2)
                (\rho_{,\mu} + \gamma_{,\mu})^2 \nonumber \\
         & & -{r^2\over 2}(1 - \mu^2)\gamma_{,r}(\rho_{,r} + \gamma_{,r})
                (\rho_{,\mu} + \gamma_{,\mu}) \nonumber \\
         & & + {1\over 4}(1 - \mu^2)\gamma_{,\mu}[r^2(\rho_{,r} +
                \gamma_{,r})^2 - (1 - \mu^2)(\rho_{,\mu} + \gamma_{,\mu})^2]
	\nonumber \\
         & & + (1 - \mu^2)e^{-2\rho}\biggl[ {1\over 4}r^4\mu
               \omega_{,r}^2 + {1\over 2}r^3(1 - \mu^2)\omega_{,r}\omega_{,\mu}
                - {1\over 4}r^2\mu(1 - \mu^2)\omega_{,\mu}^2 \nonumber \\
        & & + {1\over 2}r^4(1 - \mu^2)\gamma_{,r}\omega_{,r} 
               \omega_{,\mu} - {1\over 4}r^2(1 - \mu^2)\gamma_{,\mu}
                [r^2\omega_{,r}^2 - (1 -\mu^2)\omega_{,\mu}^2]\biggr]
	\nonumber \\
         & & -r^2[\mu - (1 - \mu^2)\gamma_{,\mu}]e^{2\sigma}4\pi
          (t^{\hat r}_{\hat r}-t^{\hat \theta}_{\hat \theta})
	{\delta (\mu) \over r}\nonumber \\ 
	& &+ r^2(1 - \mu^2)^{1/2}(1 + r\gamma_{,r})e^{2\sigma}8\pi
          t^{\hat r}_{\hat \theta}{\delta (\mu) \over r} \Biggr\}
	\label{eq:ein4}.
\end{eqnarray}
The terms containing $t^{\hat \theta}_{\hat \theta}$ and $t^{\hat r}_
{\hat \theta}$ in the above equation are identically zero for an
equatorial disk
since each particle has no $\theta$-component of momentum.  The remaining matter term in equation~(\ref{eq:ein4}), 
\begin{equation}
	-r^2[\mu - (1 - \mu^2)\gamma_{,\mu}]e^{2\sigma}4\pi
          t^{\hat r}_{\hat r}	{\delta (\mu) \over r}
\end{equation}
also does not contribute since $\gamma ,_{\mu} \vert _{\mu = 0}  $vanishes
by symmetry.

The stress-energy tensor for the matter is determined by the phase-space
distribution function $f$, which is governed by the relativistic
Vlasov equation (the matter dynamical equation).
Any distribution function of the form $f=f(E,J_z)$ is an
equilibrium solution of the Vlasov equation in axisymmetry.
Here $E$ and $J_z$ are the two constants of motion associated with the
Killing vectors $\partial/\partial t$ and $\partial/\partial\phi$.  For
our two-dimensional disk system they are
\begin{equation}
	E \equiv -{\bf p}\cdot{\partial\over\partial t}= e^{\nu_e} p^{\hat t}
	+\omega e^{\beta_e} r p^{\hat\phi},\label{eq:E}
\end{equation}
\begin{equation}
	J_z \equiv {\bf p}\cdot{\partial\over\partial \phi}=e^{\beta_e} 
  	r\,p^{\hat\phi},\label{eq:jz}
\end{equation}
where $p^{\hat \alpha}$ are the orthonormal components of the particle
4-momentum $\pbf$ in the ZAMO frame and we have defined, for convenience, the quantities,  
\begin{eqnarray}
\nu _e &\equiv & \nu (r,0),\nonumber \\
\beta_e &\equiv & \beta (r,0), \nonumber \\
\omega_e &\equiv & \omega (r,0),\quad  etc.  \label{eq:nomu}
\end{eqnarray}
Physically, $E$ is the conserved 
energy of a particle
and $J_z$ is the conserved angular momentum about the symmetry axis.

Working in two-dimensions, we define the disk stress-energy tensor for the matter, 
\begin{equation}
	t^{\hat \alpha \hat \beta}=\int f 
	p^{\hat\alpha}p^{\hat\beta}{d^2\hat p
	\over p^{\hat t}}.\label{eq:seint}
\end{equation}
Here, for a disk, we have
\begin{equation}
	p^{\hat t}=[(p^{\hat r})^2+
	(p^{\hat \phi})^2+m^2]^{1/2},\label{eq:pt}
\end{equation}
and
\begin{equation}
	d^2\hat p=dp^{\hat r}dp^{\hat \phi},\label{eq:d2p}
\end{equation}

\noindent
where $m$ is the particle mass.
For the field equations~(\ref{eq:ein1}) -- (\ref{eq:ein3}) and (\ref{eq:ein4}), 
we need
only the combinations
\begin{equation}
	t_1 \equiv t^{\hat \phi}_{\hat \phi} -t^{\hat t}_{\hat t},
\end{equation}
\begin{equation}
	t_2 \equiv t^{\hat r}_{\hat r},
\end{equation}
\begin{equation}
	t_{03}\equiv t^{\hat t}_{\hat \phi}.
\end{equation}
Using equations~(\ref{eq:seint}) -- (\ref{eq:d2p}), we get
\begin{equation}
	t_1(r) = \int d p^{\hat r}
	\int d p^{\hat\phi} {(p^{\hat r})^2 +
	2(p^{\hat\phi})^2 + m^2 \over [(p^{\hat r})^2 + 
	(p^{\hat\phi})^2 + m^2]^{1/2}}f(E,J_{z}),\label{eq:t1}
\end{equation}
\begin{equation}
	t_2(r) = \int d p^{\hat r}\int d p^{\hat\phi} {(p^{\hat r})^2 \over [(p^{\hat r})^2 
	+ (p^{\hat\phi})^2 + m^2]^{1/2}} f(E,J_{z}),\label{eq:t2}
\end{equation}
\begin{equation}
	 t_{03}(r) = \int  dp^{\hat r}
	\int d p^{\hat\phi} p^{\hat\phi} f(E,J_{z}).\label{eq:t03}
\end{equation}
At each spatial point $(r,\mu)$, the limits of integration in equations~(\ref
{eq:t1}) - (\ref{eq:t03}) will be determined by the distribution function, as
discussed in the next subsection.
\subsection{Distribution Function}\label{distfunc}

The Newtonian Kalnajs distribution function is given by
equation~(\ref{eq:apAfnt}). 
We construct a relativistic generalization by letting $E_N \rightarrow E-m$:
\begin{equation}
 f_{GR}(E,J_z) = \left\{ \begin{array}{ll}
				K\biggl [
	2\{ (b V J_z/ R_m) - ({E -m })
	\} -V^2(1+b^2)\biggr ]^{-1/2}
				& \mbox{$[\cdots]>0$}\\
				0 & \mbox{$[\cdots]\le 0$}
				\end{array}
				\right .  \label{eq:fgr}
\end{equation}
Here $V$, $b$ and $R_m$ are constants which in the Newtonian limit have
the following interpretations:  
\begin{equation}
	b \equiv {\hbox {Mean angular rotation rate }
	\over \hbox {Angular speed of a 
	circular  orbit}} \equiv 
	{\Omega \over \Omega_{circ}}
	\equiv {\Omega R_m \over V},\label{eq:b}
\end{equation}
and $R_m$ is the matter radius.
 Note that
in the Newtonian limit we have
\begin{equation}
	E-m\rightarrow E_N \quad\quad J_z \rightarrow J_{zN},\label{eq:jzntlim}
\end{equation}
where $E_N$ and $J_{zN}$ are the Newtonian energy and angular momentum
as defined by equations~(\ref{eq:apAEn}) and (\ref{eq:apAjzn}).
Thus we recover the Newtonian distribution function $f_N$, equation~
(\ref{eq:apAfnt}), from equation~(\ref{eq:fgr}) 
in the Newtonian limit.

Since the distribution function is nonzero only for positive values of the
argument of the square root, we can solve for the limits of integration for
the stress-energy integrals (eqs.~(\ref{eq:t1}) -(\ref{eq:t03})) by using the
expressions~(\ref{eq:E})  and~(\ref{eq:jz})  for $E$
and $J_{z}$ in equation~(\ref{eq:fgr}) and setting the argument of the 
square root to zero.  Thus we find the limits to be
\begin{equation}
	0<p^{\hat r} < p^{\hat r}_{max}(p^{\hat \phi},r)
\label{prlims}
\end{equation}
and
\begin{equation}
	p^{\hat \phi}_- (r) < p^{\hat\phi}<p^{\hat\phi}_+(r)
\label{pphilims}
\end{equation}
with
\begin{equation}
	p^{\hat r}_{max}(p^{\hat \phi},r) \equiv \Biggl [{a^2 + d^2 -1 
	\over 1-d^2} - (1-d^2)\biggl (p^{\hat\phi} - {ad\over 1-d^2}
	\biggr )^2 \Biggr ]^{1/2}\label{eq:prmax}
\end{equation}
and
\begin{equation}
	p^{\hat\phi}_\pm \equiv {ad\over 1-d^2} \pm {( a^2
	+ d^2 -1  ) ^{1/2}\over 1-d^2 } . \label{eq:pphipm}
\end{equation}
Here  $a$ and $d$ are functions of $r$ only and are given by
\begin{equation}
	a(r) \equiv e^{-\nu _e(r)}\biggl [1- {V^2 \over 2}(1-b^2)
	\biggr ]\label{eq:a}
\end{equation}
and
\begin{equation}
	d(r) \equiv e^{\beta _e(r) - \nu _e(r)}r\biggl [b{V\over R_m} - 
	\omega _e(r)\biggr ].\label{eq:d}
\end{equation}

Note that in \cite{st1} the distribution functions considered contained a 
parameter $E_{max}$ that was less than $m$ for bound systems of finite
extent.  The limits of integration were determined by $E\le E_{max}$.  
Here, one can verify that the limits (\ref{prlims})-(\ref{pphilims}) never
violate the condition $E\le m$.  Curiously, equality can be attained even
in the Newtonian limit: $E_N \rightarrow 0$ at the outer edge of a ``cold''
Kalnajs disk.

The total mass-energy of the system is
\begin{eqnarray}
M &=& -\int{\left(2T^\mu_\nu - \delta^\mu_\nu T\right)\xi^\nu_{(t)}
                        d^3\Sigma_\mu} \nonumber \\
        &=& \int{\left(-2T^t_t + T\right)\sqrt{-g} d^3x}.
\end{eqnarray}
Here, $\xi^\nu_{(t)}=\partial/\partial t$ is the time Killing vector.
Transforming the integrand to the ZAMO frame, and integrating out the
$\delta$-function in $\mu$, we obtain 
\begin{equation}
M= 2 \pi \int \left [t_1(r) + t_2(r) + 2\omega_e e^{-\rho_e} r t_{03}(r)\right ]
 e^{2 \sigma_e + \gamma_e} r dr\label{eq:M}.
\end{equation}
As we will see in Section~\ref{numsol}, 
this equation plays a crucial role in the
iterative solution of the combined matter and field equations.

\section {\bf Diagnostic Probes}\label{diagprobes}

There are a number of useful quantities that characterize an equilibrium
system once a solution has been obtained.
The total angular momentum is given by
\begin{eqnarray}
J &=& \int{T^\mu_\nu \xi^\nu_{(\phi)} d^3\Sigma_\mu}\nonumber \\
        &=& \int{T^t_\phi \sqrt{-g} d^3x}.
\end{eqnarray}
Transforming to the ZAMO frame, and again integrating out the delta function
in $\mu$, we get
\begin{equation}
	J= 2 \pi \int t_{03}(r)
 	e^{2 (\sigma_e + \beta_e)} r^2 dr.\label{eq:J}
\end{equation}
\noindent
The surface rest mass density is given by
\begin{equation}
	\sigma_0 (r)  = m \int  dp^{\hat r} 
	\int dp^{\hat \phi} f_{GR}(E,J_z),\label{eq:rho0}
\end{equation}
and the total rest mass by
\begin{equation}
	M_0= 2 \pi \int \sigma_0 (r) e^{2 \sigma_e + \beta_e} r dr \label{eq:M0}
\end{equation}
The binding energy of the system is defined as
\begin{equation}
	E_b\equiv M_0-M,\label{eq:eb}
\end{equation}
and can be computed from equations~(\ref{eq:M}) and (\ref{eq:M0}).

\section {\bf Scaling and Non-dimensional Units}\label{scaling}

The quantities $m$ and $M$ can be scaled out of all the above equations.
For example, we can define
\begin{eqnarray}
&\tilde p& = p/m, \quad \tilde r = r/M,\quad \tilde J = J/M^2\nonumber \\
&\tilde t^{\alpha \beta}& = t^{\alpha \beta}/M^{-1},\quad \tilde f_{GR}
= f_{GR}/M^{-1}m^{-3},\quad \tilde J_z = J_z /m M. \label{eq:scaling}
\end{eqnarray}

\noindent
With these definitions, all the previous equations can be written in tilde 
variables without $m$ or $M$ appearing.  Equivalently, the original
equations can be solved setting $m=M=1$ and scaling the final results according to equation~(\ref{eq:scaling}) to accommodate arbitrary values of $m$ and $M$.
Henceforth we will make this simplification.

In the Newtonian limit there is an additional scale freedom in that we can 
also set $R_m/M$ to one.  This is not the case for relativistic systems.

\section{\bf Numerical Solution}\label{numsol}

The numerical scheme used here is a 2-dimensional analogue of the 
procedure adopted in \cite{st1}.  In this procedure we start with initial
guesses for the metric potentials, $\rho$, $\gamma$, $\omega$, and
$\sigma$ (see section~\ref{initdata} below). We then integrate equations~(\ref{eq:t1})-(\ref{eq:t03}) to find $t_1$, $t_2$
and $t_{03}$ up to the constant factor $K$ appearing
in equation~(\ref{eq:fgr}).  This unknown factor $K$ is fixed by requiring 
that the total mass of the system, equation ~(\ref{eq:M}), satisfy $M=1$.  
Next we integrate the field equations~(\ref{eq:ein1})-(\ref{eq:ein2}) for
$\rho$, $\gamma$, and $\omega$, and then equation~(\ref{eq:ein4}) for
$\sigma$.
We then iterate
this procedure until some convergence criterion is met.

Using a combination of integral and finite differencing techniques we
solve the equations for the matter and gravitational fields.  These
equations are solved on a discrete grid in $\mu$ and $r$ on the 
computational domain  $0 \le r \le \infty$ and
$0 \le \mu \le 1$.  Unlike the Newtonian case, we cannot restrict
the computational domain to the matter interior because the
matter-independent effective source terms  $R_\rho$,
$R_\gamma$ and $R_\omega$ are nonzero in the vacuum exterior.  
Consequently we divide the radial grid into an interior and exterior domain.
Each domain is covered by a geometrically spaced grid in $r$, with the
grids joined smoothly at the cluster surface.  We use an angular grid
that is uniformly spaced in $\mu$.  The interior radial grid is
arranged to yield sufficient resolution for the core of the cluster, 
while the outer grid extends to some sufficiently large radius,
typically $3$ times the radius of the matter surface.  High resolution
of the core is crucial for obtaining numerical accuracy in highly 
centrally condensed relativistic disks.

The three elliptic
field equations~(\ref{eq:ein1}) -- (\ref{eq:ein3}) 
are solved by an integral Green's function
approach following \cite{koma}
and \cite{cook()}.  Again we make a matter-dependent, matter-independent 
split of the Green's functions depending on whether we are integrating 
the matter-dependent source terms $t_{\rho}$, $t_{\gamma}$, and $t_{\omega}$
or the matter-independent source terms $R_{\rho}$, $R_{\gamma}$ and 
$R_{\omega}$.  
Thus we write the solution of equation~(\ref{eq:ein1}) as
\begin{equation}
\rho(r,\mu) = W_{\rho}(r,\mu) + G_{\rho}(r,\mu)
\end{equation}
where we have defined
\begin{equation}
W_{\rho}(r,\mu) \equiv - \sum^{\infty}_{n=0} e^{- \gamma/2} \int_0^{\infty}
	dr^{\prime} \int ^1_0 d \mu^{\prime} r^{\prime 2} f^2_{2n} (r,
	r^{\prime}) P_{2n} (\mu) P_{2n} (\mu^{\prime}) R_{\rho}(r^{\prime},
	\mu^{\prime}),\label{eq:wrho}
\end{equation}
and
\begin{equation}
G_{\rho}(r,\mu) \equiv -{1\over 2}\sum^{\infty}_{n=0} 
	e^{- \gamma/2}P_{2n} (\mu)
	{(-1)^n(2n-1)!! \over 2^n n!}\int_0^{\infty}dr^{\prime}r^{\prime}
	f^2_{2n} (r,r^{\prime})t_{\rho}( r^{\prime},0).\label{eq:grho}
\end{equation}
Here we have used the $\delta$-function in $\mu$ to carry out the $\mu 
^{\prime}$ integration in $G_{\rho}$.
Similarly to solve equation~(\ref{eq:ein2}) we write
\begin{equation}
r \sin \theta\, \gamma(r,\mu) = W_{\gamma}(r,\mu) + G_{\gamma}(r,\mu)
\end{equation}
with
\begin{eqnarray}
W_{\gamma}(r,\mu) &\equiv &  -{2 \over \pi} \sum^{\infty}_{n=1} e^{-\gamma/2}
	\int_0^{\infty} dr^{\prime} \int^1_0 d 
	\mu^{\prime} r^{\prime 2}
	f^1_{2n-1} (r, r^{\prime}) {1 \over 2n-1} 
	\sin(2n-1) \theta\nonumber \\
	& &\times \sin(2n-1) \theta^{\prime} 
	R_{\gamma} (r^{\prime}, \mu^{\prime}),\label{eq:wgam}
\end{eqnarray}
and
\begin{equation}
G_{\gamma}(r,\mu) \equiv {1\over \pi}\sum^{\infty}_{n=1} e^{-\gamma/2}
	{\sin(2n-1) \theta (-1)^n\over (2n-1)}\int_0^{\infty} dr^{\prime}
	f^1_{2n-1} (r, r^{\prime})t_{\gamma}(r^{\prime},0).\label{eq:ggam}
\end{equation}
Finally, to solve equation~(\ref{eq:ein3}) we write
\begin{equation}
r\sin\theta\,\omega (r,\mu) = W_{\omega}(r,\mu) + G_{\omega}(r,\mu)
\end{equation}
with
\begin{eqnarray}
W_{\omega}(r,\mu)&\equiv&  - \sum^{\infty}_{n=1} e^{(2\rho- \gamma)/2}
	\int^{\infty}_{0} dr^{\prime} \int^{1}_{0}  
	d \mu^{\prime} r^{\prime 3} \sin
	\theta^{\prime} f^2_{2n-1} (r, r^{\prime}) 	
	{1 \over 2n(2n-1)}\nonumber \\
	& &\times P^1_{2n-1} (\mu) P^1_{2n-1} 
	(\mu^{\prime})R_{\omega}(r^{\prime},
	\mu^{\prime}),\label{eq:womg} 
\end{eqnarray}
and
\begin{equation}
G_{\omega}(r,\mu) \equiv -{1\over 2}\sum^{\infty}_{n=1} e^{(2\rho
 	- \gamma)/2}{P^1_{2n-1} (\mu)\over 2n(2n-1)}{(-1)^n (2n-1)!!
	\over 2^n (n-1)!}\int^{\infty}_{0} dr^{\prime} r^{\prime 2}
	f^2_{2n-1} (r, r^{\prime})t_{\omega}(r^{\prime},0).\label{eq:gomg}
\end{equation}
Here
\begin{equation}
f^1_n (r, r^{\prime}) = \left \{ \begin{array}{ll}
 	(r^{\prime} /r)^n, & \mbox {for  $r^{\prime}/r \leq 1$}\\
	(r/r^{\prime})^n, & \mbox {for $r^{\prime}/r > 1$}
	\end{array}
	\right .
\end{equation}
and
\begin{equation}
f^2_n (r, r^{\prime}) = \left \{ \begin{array}{ll}
		 (1/r)(r^{\prime}/r)^n, & \mbox {for $r^{\prime}/r\leq 1$,}\\
	(1/r^{\prime})(r/r^{\prime})^n, & \mbox {for $r^{\prime}/r > 1$}
	\end{array}
	\right .
\end{equation}

Among the advantages of this Green's function approach for solving the
elliptic field equations is that the asymptotic conditions on $\rho$,
$\gamma$ and $\omega$ are imposed automatically. That is,
$\rho \sim \Order(1/{r})$, $\gamma \sim \Order(1/{r}^2)$, and
$\omega \sim \Order(1/{r}^3)$ for large ${r}$.
To improve the accuracy of the angular integrations, we use the identities
in equations (34) -- (38) of \cite{cook()}.

It should be noted that in the integration of equation~(\ref{eq:gomg}), the calculation
of $t_{\omega}(r,0)$ at $r=0$ requires special care.  It can be seen
from equations~(\ref{eq:t03}), (\ref{eq:prmax}) and (\ref
{eq:pphipm}) that $t_{03}(r=0) = 0$, while regularity conditions near the
axis imply that $t_{03} \sim r $ near $r=0$.  Thus, at $r=0$, the calculation
of the quantity $t_{03} /r$ that appears in equation~(\ref{eq:tomg}) is
done by computing the quantity $dt_{03}/dr \vert _{r=0}$ analytically.
We do this by taking a derivative of equation~(\ref{eq:t03}) directly. For
any $r$ this gives
\begin{eqnarray}
dt_{03}/dr  &=& {d \over dr}
\biggl [\int^{p_+^{\hat\phi}(r)}_{p_-^{\hat\phi}(r)} 
	d p^{\hat\phi} p^{\hat\phi} \int_0^
	{p^{\hat r}_{max}(r,p^{\hat \phi})}
	  dp^{\hat r}
	f_{GR}(E,J_{z})\biggr ]\\ \nonumber
&=& \int^{p_+^{\hat\phi}(r)}_{p_-^{\hat\phi}(r)} dp^{\hat\phi} p^{\hat\phi}
	{\partial I(p^{\hat\phi},r) \over \partial r} \\ 
&\quad& + p_+^{\hat\phi}(r)
	I(p_+^{\hat\phi}(r),r){dp_+^{\hat\phi}(r)\over dr}
	- p_-^{\hat\phi}(r)I(p_-^{\hat\phi}(r),r)
{dp_-^{\hat\phi}(r)\over  dr}.\label{dt03dr}
\end{eqnarray}
Here $p_{\pm}^{\hat\phi}(r)$ and $p^{\hat r}_{max}(r,p^{\hat \phi})$ are
defined in equations~(\ref{eq:pphipm}) and (\ref{eq:prmax}) and  
 we have defined $I(p^{\hat\phi},r)$ to be
\begin{equation}
I(p^{\hat\phi},r) = \int_0^
	{p^{\hat r}_{max}(r,p^{\hat \phi})}
	  dp^{\hat r}
	f_{GR}(E,J_{z}).\label{igr}
\end{equation}
At $r=0$ it can be shown that the first term in equation~(\ref{dt03dr})
vanishes by symmetry.  The other two terms involve the integral $I(p^{\hat\phi},r)$ evaluated at $p^{\hat\phi}=p_{\pm}^{\hat\phi}(0)$ and $r=0$.  From
equation~(\ref{igr}) and the form of the distribution function $f_{GR}$ 
(equation~(\ref{eq:fgr})) we see that this integrand  is divergent and the
limits collapse to zero at
$p^{\hat\phi}=p_{\pm}^{\hat\phi}(0)$.  However, near $p^{\hat\phi}=p_{\pm}^{\hat\phi}(0)$ we can expand the integrand (and upper limit) to get an estimate of the value of the
integral for use in the calculation of $dt_{03}/dr \vert _{r=0}$.  This
yields
\begin{equation}
dt_{03}/dr \vert _{r=0} = \pi\biggl [ \sqrt{a(r)(a^2(r)-1)}\,\,a(r)\,d_{,r}(r)
e^{-\nu_e/2}\biggr ]_{r=0}. \label{dt03drfinal}
\end{equation}
Here $a(r)$ and $d(r)$ are functions defined by equations~(\ref{eq:a})
and (\ref{eq:d}).

Equation~(\ref{eq:ein4}) for $\sigma$ is solved by integrating 
the linear ODE from the
pole ($\mu = 1$) to
the equator with the initial condition that
\begin{equation}
\sigma = {\gamma - \rho \over 2} \quad\hbox{at}\quad \mu = 1,\label{eq:sigbc}
\end{equation}
which arises from the requirement of local flatness on the coordinate axis.
The derivatives of $\rho$, $\gamma$ and $\omega$
appearing in the matter-independent source terms $R_\rho$, $R_\gamma$, 
$R_\omega$ and
the right-hand side of equation~(\ref{eq:ein4}) are evaluated by finite 
differencing.

Since we can evaluate the integrands in equations~(\ref{eq:t1}) -- (\ref{eq:t03})
at any values of
$p^{\hat r}$ and $p^{\hat\phi}$, we carry out the quadrature over each variable
by Romberg integration (\cite{press}).

For a convergent solution, we require the maximum fractional change in
all four metric functions to be less than $1 \times 10^{-3}$
on successive iterations. This typically takes about 20 iterations when
we start with a Newtonian initial guess (Section~\ref{initdata}).  A significant number
of iterations can be saved during a sequence of calculations in which the
heating parameter $b$ is varied if we use the previous solution as the initial
guess for the next value of $b$.

For disks in the relativistic region, 600 radial zones, 20 angular zones, and
20 Legendre polynomials are adequate.

\section{\bf Initial Data}\label{initdata}

As an initial guess for the metric potentials $\rho(r,\mu)$, $\gamma(r,\mu)$, $\omega(r,\mu)$,
$\sigma(r,\mu)$   we used their Newtonian limits, given in terms of the Newtonian
potential $\Phi_N$, by 
\begin{eqnarray}
\rho(r,\mu) &\rightarrow & 2\Phi _N (r,\mu),\nonumber \\
\gamma(r,\mu) &\rightarrow & 0, \nonumber \\
\sigma(r,\mu) &\rightarrow & - \Phi _N (r,\mu), \nonumber \\
\omega(r,\mu) &\rightarrow & 0. \label{eq:newpots}
\end{eqnarray}
\cite {firstNewt()} gives an expression 
for  the Newtonian potential of an oblate homogeneous spheroid.  By taking
the limit as the eccentricity  $e\rightarrow 1$, we obtain the Newtonian
potential $\Phi _N(r,\mu)$ for all points both on and off the disk. Outside 
the matter, we find
\begin{equation}
\Phi _N(r,\mu) = 
	-{3 M \over 2 R_m}\biggl [B - {q_s \over 2(1-x^2)^{1/2}} 
	\bigg\lgroup (1-3\mu^2)(B(1-x^2)^{1/2} -x) + (1-\mu^2)x^3\bigg \rgroup \biggr ].\label{eq:Newtphi1}
\end{equation}
where $q_s \equiv r^2/R_m^2$ , $x  = \sqrt{p}$, and $B =\arcsin {x}$
with $p$ defined by the quadratic equation
\begin{equation}
q_s(1-\mu^2)p^2 - (q_s +1)p +1 = 0. \label{eq:pquad}
\end{equation}
Inside the matter, we have
\begin{equation}
	\Phi _N(r,\mu) = 
-{3\pi  M \over 4 R_m}(1-{q_s \over 2}).  \label{eq:Newtphi2}
\end{equation}

\section{\bf Numerical Results}\label{numresults}
\begin{figure}[!htb]
\begin{center}
\begin{picture}(325,200)
\put(0,0){\epsfxsize=5in\epsffile{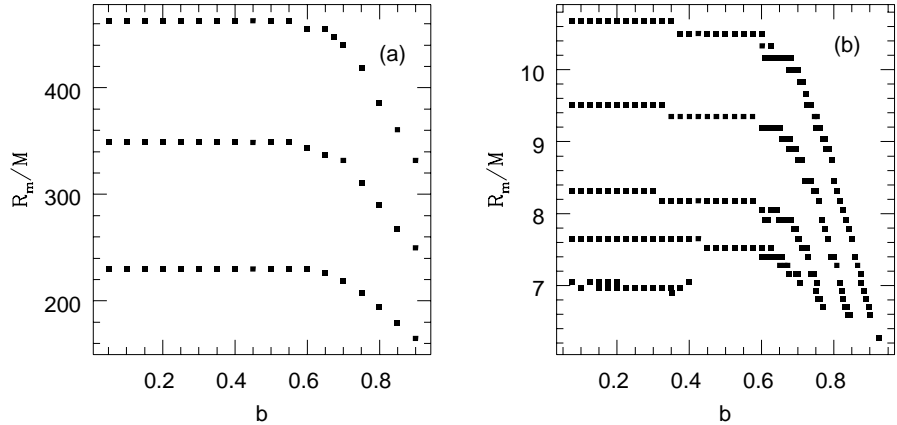}}
\end{picture}
\end{center}

\caption{ Convergent runs for some representative values of
$R_m/M$.  (a) Runs in the Newtonian region.  (b) Runs in the more
relativistic region.
\label{fig:bpvsrs}
}

\end{figure}

Our region of investigation spans values of the heating parameter
in the range $.05 \le b \le .95$ for values of $R_m/M$ ranging from 
$R_m/M = 455$ down to $R_m/M = 6.26$.
Figure~\ref{fig:bpvsrs} (a) shows the convergence of 
some representative runs in the Newtonian 
region. As we can see, as $b$ increases, $R_m/M$ of the solution
decreases.  This is also evident for the more relativistic cases
shown in Figure~\ref{fig:bpvsrs} (b). 
\noindent

\begin{figure}[!htb]
\begin{center}
\begin{picture}(325,200)
\put(0,0){\epsfxsize=5in\epsffile{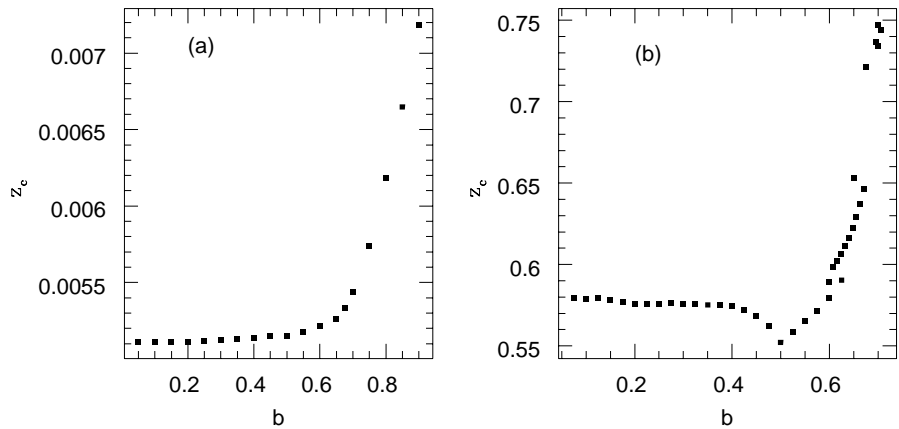}}
\end{picture}

\end{center}

\caption{ Central redshift ($z_c$) versus heating parameter 
$b$ for our most Newtonian cluster, (a)\,\,$R_m / M = 455$ and 
a representative relativistic cluster,\,\,(b)\,\,$R_m / M = 8.84$ .
\label{fig:bpvszc}
}
\end{figure}

To better understand our solutions it is helpful to look at how the
central redshift, $z_c$, varies with the heating parameter, $b$.
For the Newtonian Kalnajs disk given in Appendix A, we can calculate 
the central
redshift, $z_c$ analytically from the Newtonian gravitational 
potential as follows:
\begin{eqnarray}
z_c &\equiv& \left [ -g_{tt}(0,0) \right ] ^{-1/2} -1 \\ \nonumber
	&=& \left [ e^{\gamma(0,0) +\rho(0,0)}\right ]^{-1/2} -1.
\label{zcgen}
\end{eqnarray}
Now use equations~(\ref{eq:newpots})
to get
\begin{equation}
z_c^{Newtonian} = -\phi _{N} (0,0) = {3\pi M \over 4 R}. \label{zcnewt}
\end{equation}
This quantity is independent of the heating parameter $b$ and thus
is a useful diagnostic.  Although analytic equilibrium solutions exist 
in the Newtonian case even in the unstable region ($b \ge
.816$) (see Appendix A), it is not obvious that our iterative method
will converge to them.  Instabilities may be signaled by failure
to converge, or by causing convergence to a solution that is far from
the true Kalnajs solution.  Thus by plotting the central redshift versus the
heating parameter we can hope to see  where the solution we find differs from 
the true Newtonian solution. As can be seen in Figure~\ref{fig:bpvszc} (a), for
a Newtonian cluster with $R_m/M = 455$, our code finds the Kalnajs equilibrium
solutions for $b \le .6$.  Since the stability region
is given by $0\le b < .816$,  the change in the central redshift
does not serve as a sharp indicator of the transition to the unstable
region.

Similar behavior can be observed in our relativistic clusters.
Figure~\ref{fig:bpvszc} (b) shows an example for $R_m/M = 8.84$. Regions in which $z_c$ is not constant
may represent true general relativistic equilibria, but most likely
these clusters are near or inside the general relativistic unstable
region.  To fully pin down this region requires a dynamical stability
analysis, which can probably only be done by numerical evolution.
\noindent

The most relativistic solution we have found has an $R_m/M$ of $6.97$. By
comparing this disk to the Kerr geometry we can get an idea of how relativistic
this model is.  Using the proper circumference of the disk we find that our
most relativistic cluster corresponds to an $R/M$ for Kerr of $8.09$.  Table~\ref{table1} gives some properties of the representative models
discussed here.
\begin{deluxetable}{cccccc}
\tablecaption{Properties of some representative solutions.}
\tablewidth{0pt}
\tablehead{
\colhead{$R_m/M$}           & \colhead{$b$}      &
\colhead{$z_c$}          & \colhead{$g_{tt}^{max\,\,}$\tablenotemark{a}}  &
\colhead{$E_b$\tablenotemark{b}}          & \colhead{$J$\tablenotemark{c}}}
\startdata
455 & 0.6 & 0.0051 & -0.9897 & -0.0032 & 4.2304\nl 
6.97 & 0.1 & .9556 & -0.2614 & \nodata \tablenotemark{d} & 0.0693\nl
\enddata
\tablenotetext{a}{$g_{tt}^{max}$ is the maximum value of $g_{tt}$.}
\tablenotetext{b}{$E_b$ is in units of $M_0$.}
\tablenotetext{c}{$J$ is in units of $M^2$}
\tablenotetext{d}{For the relativistic disk here our value of $E_b$
is unreliable because of numerical errors. }
\label{table1}
\end{deluxetable}

\begin{figure}[!htb]
\begin{center}
\begin{picture}(250,250)
\put(0,0){\epsfxsize=3.5in\epsffile{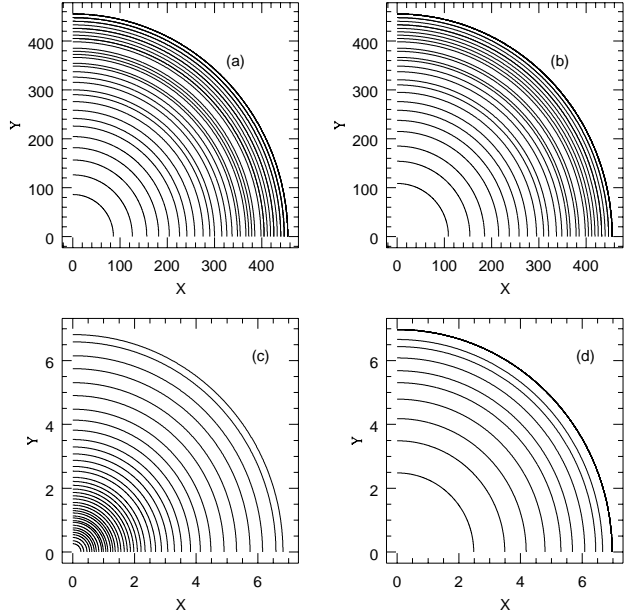}}
\end{picture}

\end{center}

\caption{Surface density contour plots: 
(a)\,\,Our most Newtonian cluster with $R_m/M = 455$. and \,(b)\, its analytic
analogue.\,(c) Our most relativistic cluster with
$R_m/M = 6.97$. (d) The analytic Newtonian version of this cluster ($R_m/M = 6.97$).
\label{fig:contours}
}
\end{figure}

In Figure~\ref{fig:contours} we show contour plots of the surface density for two examples of
our equilibrium disks.  Frame (a) shows our most
Newtonian cluster ($R_m/M=455$) with $b= .6$.  In Newtonian
theory these disks have homogeneous volume density. 
Frame (b)
shows the same plot for the analytic version of the same cluster.
In contrast, Frames (c) and (d) show our most relativistic
cluster ($R_m/M=6.97$) with $b = .1$ and its Newtonian
counterpart respectively.  The relativistic disks are in general more
centrally condensed than their Newtonian cousins.

In previous work (\cite{st19851992b}) on
polytropic axisymmetric clusters we found that a maximum
in the binding energy along an equilibrium sequence marked the onset
of a dynamical instability.  Since, for the Newtonian Kalnajs disks here 
the potential, (equation~(\ref{eq:apAphi})) and surface matter density, 
(equation~(\ref{eq:apAsig})) are not functions of the heating parameter, the binding
energy is also not a  function of $b$. Thus it cannot be 
used to analyze stability.  However, in the general relativistic
case we expect this degeneracy to be lifted.  We do indeed find that the 
binding energy does depend on $b$.  However, because of numerical errors, we 
are  not able to
reliably find the turning point in the binding energy that would  signal
the onset of instability. Again, a firm limit on the
stability region must wait for a full dynamical evolution code.
\acknowledgements
We thank Mark Scheel and Greg Cook for useful discussions.  We also
thank Greg Cook for generously sharing computer code with us.  This work
has been supported in part by National Science Foundation grant number
PHY 94-08378 at Cornell University, and by National Science Foundation Grant AST 96-18524 and
     NASA Grant NAG 5-3420 at the University of Illinois at Urbana-Champaign.
\appendix
\section{Newtonian Kalnajs Disks}\label{NKD}

A Newtonian Kalnajs disk can be obtained by ``flattening'' 
 a homogeneous oblate spheroid, i.e., by letting the eccentricity $e\rightarrow
 1$ (\cite{bnt}).
This flattening yields a disk with a surface density given by
\begin{equation}
	\Sigma(R) = \Sigma_c (1 - R^2/R_m^2),\label{eq:apAsig}
\end{equation}
where $\Sigma_c$ is the central density, $R_m$ is the radius of the disk (matter
radius) and $R$ is the radius in the disk plane defined by
\begin{equation}
	R \equiv r\,\sin\theta	.
\end{equation}
When all the particles are in circular orbits with uniform angular velocity
$\Omega_{circ}$, then $\Omega_{circ}$ is related to the central density by
the same relation as for Maclaurin spheroids.

  \cite{kal()} developed a family of equilibrium disks, all of
which have the above surface density.  Each member of this family is 
characterized by the parameter $b$ defined in equation~(\ref{eq:b}).

These disks have velocity dispersions that are governed by the tunable
parameter $b$:
\begin{equation}
	\overline{\left ( v^{\hat \phi}\right )^2} -  \left (\overline{ v^{\hat \phi}}
	\right )^2 =
	\overline {\left (v^{\hat R}\right )^2}=
	{1\over 3}V^2(1-b^2)(1-R^2/R_m^2).\label{eq:apAvdis}
\end{equation}
From the above equation we can see that when $b\approx 1$ 
($\Omega R_m \approx V$)
these disks are ``cold'', i.e. have very little thermal motion.  
This corresponds physically
to a system in which all particles move on nearly circular orbits.
On the other hand, when $b\ll 1$ ($\Omega R_m \ll V$) we have ``hot'' systems in which
most of the support against self-gravity comes from the random motions
given in equation~(\ref{eq:apAvdis}) (\cite{bnt}). 

A distribution functions that generates this family is
\begin{equation}
 f_N(E_N,J_{zN}) = \left\{ \begin{array}{ll}
				K_N\biggl [
	2\{ (b V J_{zN}/ R_m) - {E_N}
	\} -V^2(1+b^2)\biggr ]^{-1/2}
				& \mbox{$[\cdots]>0$}\\
				0 & \mbox{$[\cdots]\le 0$}
				\end{array}
				\right .  \label{eq:apAfnt}
\end{equation}
Here $K_N$ is a constant that can be determined by fixing the total
mass of the system, 
 $E_N$ is the Newtonian energy, and $J_{zN}$ is the Newtonian
 angular momentum
about the symmetry ($z$) axis.  These quantities, with $m=1$, are given by
\begin{equation}
	E_N = {(p^{\hat R})^2\over 2} + {(p^{\hat \phi})^2\over 2} + \phi_N(R)\label
	{eq:apAEn}
\end{equation}
and 
\begin{equation}
	J_{zN} = Rp^{\hat \phi}.\label{eq:apAjzn}
\end{equation}
The Newtonian potential, $\phi_N(R)$, 
in the disk plane is 
\begin{equation}
	\phi_N(R) \equiv \Phi _N (r,0) =  -V^2(1-{R^2\over 2R_m^2})
\label{eq:apAphi}.
\label{eq:apAphin}
\end{equation}

Using the linearized collisionless Boltzmann equation (linearized Vlasov
equation) \cite{kal()} analyzed the stability of this family of disks.  He 
found that restricting the heating parameter, $b$, to be less than $0.816$
produced disk systems that are stable against all axisymmetric disturbances,
i.e. disks with $0\le b < .816$ are stable against ring formation.


\begin{thebibliography}{}

\bibitem[Abrahams {\it et.al} (1994)]{nocolldisk}
Abrahams A. M., Shapiro S. L., Teukolsky S. A., 1994, {\it Phys. Rev. D.},
{\bf 50}, 7282.
\bibitem[Bardeen 1970]{bar}
Bardeen, J. M. 1970, {\it ApJ}, {\bf 162}, 71.
\bibitem[Bardeen \& Wagoner 1969]{olddisk2}
Bardeen, J. M., \& Wagoner, R. V. 1969, {\it ApJ}, {\bf 158}, L65.
\bibitem[Binney \& Tremaine 1987]{bnt}
Binney, J., \& Tremaine, S. 1987, {\it Galactic Dynamics} (Princeton: Princeton
Univ. Press), Ch.4
\bibitem[Cook \etal 1992]{cook}
Cook, G. B., Shapiro, S. L., \& Teukolsky, S. A. 1992, {\it ApJ}, {\bf 398},
 203.
\bibitem[Kalnajs 1972]{kal}
Kalnajs, A. J. 1972, {\it ApJ}, {\bf 175}, 63-76.
\bibitem[Komatsu \etal (1989)]{koma}
Komatsu, H., Eriguchi, Y., \& Hachisu, I. 1989, {\it MNRAS}, {\bf 237}, 355.
\bibitem[Fridman \& Polyachenko 1984]{frid}
Fridman, A. M., \& Polyachenko, V. L. 1984, {\it Physics of Gravitating
Systems} (New York:  Springer).
\bibitem[Lightman \etal 1975]{zamo}
Lightman, A. P., Press, W. H., Price, R. H., \& Teukolsky, S. A. 1975,
{\it Problem Book in Relativity and Gravitation} (Princeton:  Princeton Univ.
Press)
\bibitem[Meinel 1997]{anfluiddisk}
Meinel R. Pre-Print, gr-qc/9703077.
\bibitem[Mihalas 1969]{firstNewt}
Mihalas, D. 1968, {\it Galactic Astronomy} (Freeman, San Francisco).
\bibitem[Morgan \& Morgan 1969]{olddisk1}
Morgan T. \& Morgan L. 1969, {\it Phys. Rev.}, {\bf 183}, 1097.
\bibitem[Nakamur \etal 1988]{Newt}
Nakamura, T., Shapiro, S. L., \& Teukolsky, S. A. 1988, {\it Phys. Rev. D.},
{\bf 38}, 2972.
\bibitem[Press \etal 1992]{press}
Press, W. H., Teukolsky, S. A., Vetterling,
W. T., \& Flannery, B. P. 1992, {\it Numerical Recipes in Fortran:
The Art of Scientific Computing} (2nd ed.; Cambridge: 
Cambridge University Press).
\bibitem[Salpeter \& Wagoner 1971]{olddisk3}
Salpeter, E.E. \& Wagoner, R. V. 1971, {\it ApJ}, {\bf 164}, 557-567.
\bibitem[Shapiro \& Teukolsky 1985]{st1985}
Shapiro, S. L., \& Teukolsky, S. A. 1985, {\it ApJ}, {\bf 298},58.
\bibitem[Shapiro \& Teukolsky 1992a]{st2}
Shapiro, S. L., \& Teukolsky, S. A. 1992a, {\it ApJ}, {\bf 388}, 287.
\bibitem[Shapiro \& Teukolsky 1992b]{st1992b}
Shapiro, S. L., \& Teukolsky, S. A. 1992b, 
{\it Phil. Trans. R. Soc. Lond.A,}, {\bf 340},365.
\bibitem[Paper I]{st1}
Shapiro, S. L. \& Teukolsky,S.A. 1993, {\it ApJ}, {\bf 419}, 636-647.
\bibitem[Bardeen \& Wagoner (1969)]{olddisk2()}
\bibitem[Kalnajs (1972)]{kal()}
\bibitem[Cook \etal (1992)]{cook()}
\bibitem[Meinel (1997)]{anfluiddisk()}
\bibitem[Mihalas (1969)]{firstNewt()}
\bibitem[Morgan \& Morgan (1969)]{olddisk1()}
\bibitem[Salpeter \& Wagoner (1971)]{olddisk3()}
\bibitem[Shapiro \& Teukolsky 1985,1992b]{st19851992b}

\end{thebibliography}
\end{document}